\documentclass{aa}

\usepackage{graphicx}
\usepackage{txfonts}
\usepackage{natbib}
\bibpunct{(}{)}{;}{a}{}{,}

\begin{document}

\title{Dynamics of magnetic flux tubes in close binary stars}
\subtitle{II.~Nonlinear evolution and surface distributions}
\titlerunning{Flux tubes in close binaries II.}

\author{V.~Holzwarth\inst{1,2} \and M.~Sch\"ussler\inst{1}}
\authorrunning{Holzwarth \& Sch\"ussler}

\institute{Max-Planck-Institut f\"ur Aeronomie, Max-Planck-Str.~2,
37191 Katlenburg-Lindau, Germany \\
\email{schuessler@linmpi.mpg.de}
\and 
School of Physics and Astronomy, University of St.~Andrews, North
Haugh, St.~Andrews KY16 9SS, UK \\
\email{vrh1@st-andrews.ac.uk}}

\date{Received gestern; accepted morgen}

\abstract{
Observations of magnetically active close binaries with orbital periods
of a few days reveal the existence of starspots at preferred longitudes
(with respect to the direction of the companion star).
We numerically investigate the non-linear dynamics and evolution of 
magnetic flux tubes in the convection zone of a fast-rotating component
of a close binary system and explore whether the tidal effects are able
to generate non-uniformities in the surface distribution of erupting 
flux tubes. 
Assuming a synchronised system with a rotation period of two days and
consisting of two solar-type components, both the tidal force and the
deviation of the stellar structure from spherical shape are considered
in lowest-order perturbation theory.
The magnetic field is initially stored in the form of toroidal magnetic
flux rings within the stably stratified overshoot region beneath the
convection zone.
Once the field has grown sufficiently strong, instabilities initiate
the formation of rising flux loops, which rise through the convection
zone and emerge at the stellar surface.
We find that although the magnitude of tidal effects is rather small,
they nevertheless lead to the formation of clusters of flux tube
eruptions at preferred longitudes on opposite sides of the star, which
result from the cumulative and resonant character of the action of
tidal effects on rising flux tubes.
The longitude distribution of the clusters depends on the initial
parameters of flux tubes in the overshoot region like magnetic field
strength and latitude, implying that there is no globally unique
preferred longitude along a fixed direction.

}

\maketitle

\section{Introduction}
\label{intro}
Observational results indicate that in close, fast-rotating binaries 
like RS CVn and BY Dra systems magnetic activity is usually much more 
vigorous than in the case of the Sun \citep[see, e.g.,][and references 
therein]{1993A&AS..100..173S}.
Data sets covering time periods much longer than the rotation period of
a system reveal non-uniform longitudinal spot distributions (with
respect to the direction of the companion star), which indicate the
existence of long-lasting spot conglomerations at \emph{preferred
longitudes} where starspots persist longer or flux eruption is more
frequent than at other surface regions \citep{1995ApJS...97..513H,
1996A&A...314..153J}.
The preferred longitudes are frequently about $180\degr$ apart on
opposite sides of the stellar hemisphere
\citep[e.g.,][]{1998A&A...336L..25B, 1998A&A...340..437B} and their
dependence on rotation period, i.e., essentially on the distance
between the two components, suggests that they are related to the
proximity of the companion star \citep{1995AJ....109.2169H}.

In analogy to the `solar paradigm' for magnetic activity, it is assumed
that starspots at the stellar surface are generated by erupting
magnetic flux tubes which originate from the lower part of the outer
convection zone of the star.
By using non-linear numerical simulations we investigate the influence
of tidal effects on the dynamics and evolution of magnetic flux tubes
inside the convection zone until they erupt at the stellar surface.
Particular attention is paid to the non-uniformities in the resulting
surface distributions of erupting flux tubes and their dependence on
initial conditions.
This work complements our previous studies of the equilibrium and
linear stability properties of magnetic flux tubes in the overshoot
region at the lower boundary of the convection zone \citep[hereafter
HS{\it I}]{2000AN....321..175H,hsi}.

Sect.~\ref{moas} contains a brief summary of the magnetic flux tube 
model and the analytical approximation of the close binary model.
In Sect.~\ref{simu} we explain the accomplishment of the numerical 
simulations as well as the implementation of initial conditions.
Sect.~\ref{resu} shows the resulting latitudinal, longitudinal and 
surface distributions of erupting flux tubes in the case of a binary
systems with an orbital period of two days as well as a study 
concerning the dependence of non-uniformities on the orbital period.
Sect.~\ref{disc} contains a discussion of our results and
Sect.~\ref{conc} gives our conclusions.

\section{Model assumptions}
\label{moas}

\subsection{The flux tube model}
\label{ftmo}
We study the spot distribution on active binary stars by applying the
`solar paradigm' for magnetic activity, assuming that starspots are due
to erupting magnetic flux tubes \citep[e.g.,][]{1996A&A...314..503S}.
In this model, the magnetic field is stored in the form of toroidal
flux tubes inside the subadiabatic overshoot-region below the
convection zone at the interface to the radiative core.
Once a critical magnetic field strength is exceeded, perturbations
initiate the onset of an undulatory (Parker-type) instability and the
growth of flux loops, which rise through the convection zone and lead
to the formation of bipolar active regions and starspots upon emergence
at the stellar surface.

The numerical code we have used for the non-linear simulations is
explicated in \citet{1986A&A...166..291M} and
\citet{1995ApJ...441..886C}.
All calculations are carried out in the framework of the `thin flux
tube approximation' \citep{1981A&A...102..129S} describing the
evolution of an isolated magnetic flux tube embedded in a field-free
environment.
The surrounding convection zone is treated as an ideal plasma with
vanishing viscosity and infinite conductivity, implying the concept of
frozen-in magnetic fields.
The dynamical interaction between the flux tube and its environment is
taken into account by an hydrodynamic drag force acting upon a tube
with circular cross-section which moves relative to its environment.
The radius of the cross-section is determined according to the amount
of magnetic flux and the lateral balance between the external gas
pressure and the internal total (gas and magnetic) pressure.
Since the radiative time scale in the convection zone typically exceeds
the rising time of flux tubes, we assume the adiabatic evolution of
flux tubes.

\subsection{The binary model}
\label{bimo}
We consider a detached binary system with two solar-type components,
which move on circular orbits with an orbital period of two days.
For the active star we assume synchronised, solid-body rotation, with
the spin axis orientated perpendicular to the orbital plane.
Its internal structure is described by a tidally deformed solar model, 
whereas the companion star is considered as a point mass.
The model includes both the tidal force and the deviation of the
stellar structure from spherical shape, which are sufficiently small to
be treated in lowest-order perturbation theory (see HS{\it I},
Sect.~2.2).

Using spherical coordinates (radius $r$, azimuth $\phi$, latitude 
$\lambda$; $\phi\!=\!0$ is the direction toward the companion star), 
the dominating term of, e.g., the azimuthal component of the effective 
gravitation, $\vec{g}_\mathrm{eff}$, can be written as 
\begin{equation}
\frac{g_{\mathrm{eff},\phi}}{g_\star}
\sim 
\epsilon^3
q
\sin 2 \phi
\ ,
\label{grav}
\end{equation}
where $g_\star$ is the gravitation in the unperturbed case, $q$ is the 
mass ratio between the two components, $\epsilon= (r/a)$ the expansion 
parameter of the approximation and $a$ the distance between the two 
components.
For a weakly deformed star it can be assumed that the value of any 
stellar quantity, $f$, (pressure, density,\ldots) is still only a 
function of the effective potential, $\Psi_\mathrm{eff}= \Psi_\star + 
\Psi_\mathrm{tide}$, where $\Psi_\star$ is the gravitational potential
of the star and $\Psi_\mathrm{tide}$ the potential of the tidal
perturbation [see Eq.~(2) in HS{\it I}].
Following Eq.~(5) in HS{\it I}, the (Eulerian) perturbation of the
stellar structure, $\Delta f$, can be written as
\begin{equation}
\frac{\Delta f}{f_\star}
\propto
\frac{\Psi_\mathrm{tide}}{\Psi_\star}
\sim
\epsilon^3
q
\frac{r}{H_{f}}
\cos 2 \phi
\ ,
\label{quanvar}
\end{equation}
where the undisturbed value, $f_\star$, and its local scale height, 
$H_f$, are determined from an unperturbed stellar model.
The leading orders of the tidal effects given in Eqs.~(\ref{grav}) and 
(\ref{quanvar}), respectively, both show a $\pi$-periodicity in the 
azimuthal direction $\phi$.

Using Kepler's third law, the order of magnitude of tidal effects
becomes
\begin{equation}
\epsilon^3 q
\sim
10^{-2}
\frac{q}{1+q}
\left( \frac{r}{R_{\sun}} \right)^3
\left( \frac{M_\star}{M_{\sun}} \right)^{-1}
\left( \frac{T}{\,\mathrm{d}} \right)^{-2}
\ ,
\label{order}
\end{equation}
while owing to the ratio $r/H_f$ in Eq.~(\ref{quanvar}), the
perturbation of the stellar structure also depends on the scale height
of the considered quantity.
It is enhanced if $f$ changes in a relative small stellar layer, like
the superadiabaticity, $\delta= \nabla - \nabla_\mathrm{ad}$, inside
the thin overshoot region below the convection zone.
For the system considered here ($T= 2\,\mathrm{d}, M_\star= M_{\sun}, 
q= 1$ and $r\le R_{\sun}$), Eq.~(\ref{order}) yields $\epsilon^3 q\sim
10^{-3}$ (with $a\sim 8\,R_{\sun}$).

\section{Simulations}
\label{simu}
\begin{figure}
\centerline{\includegraphics[width=.5\hsize]{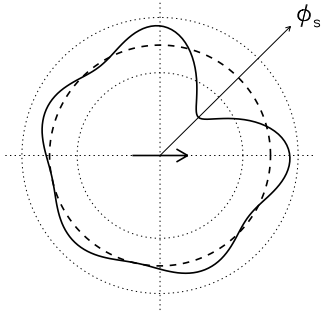}}
\caption{
Schematic illustration of the initial `localised perturbation' of a 
flux ring (\emph{dashed}) resulting from the in-phase superposition of 
sinusoidal wave modes with $m= 1\ldots5$.
The perturbation is largest around the longitude $\phi_\mathrm{s}$.
The central arrow indicates the direction toward the companion star.
}
\label{excite.pic}
\end{figure}

\begin{figure}
\includegraphics[width=.49\hsize]{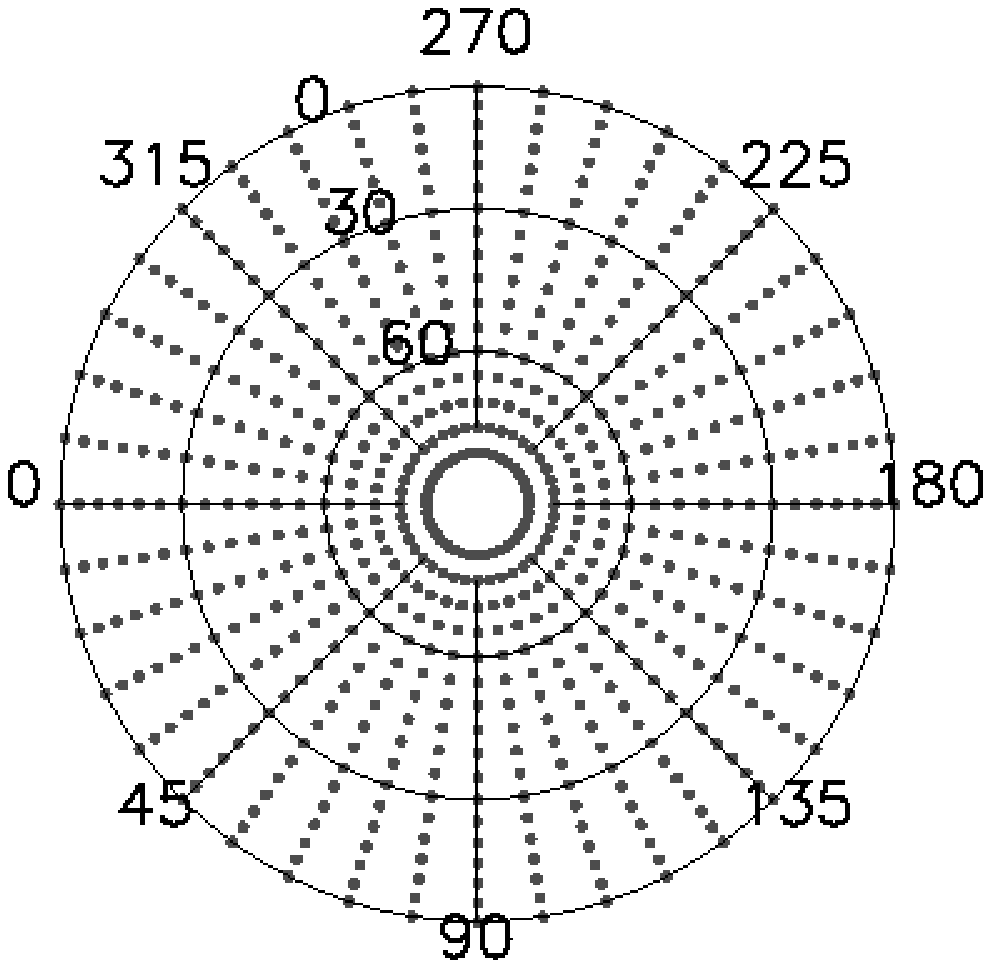}
\includegraphics[width=.49\hsize]{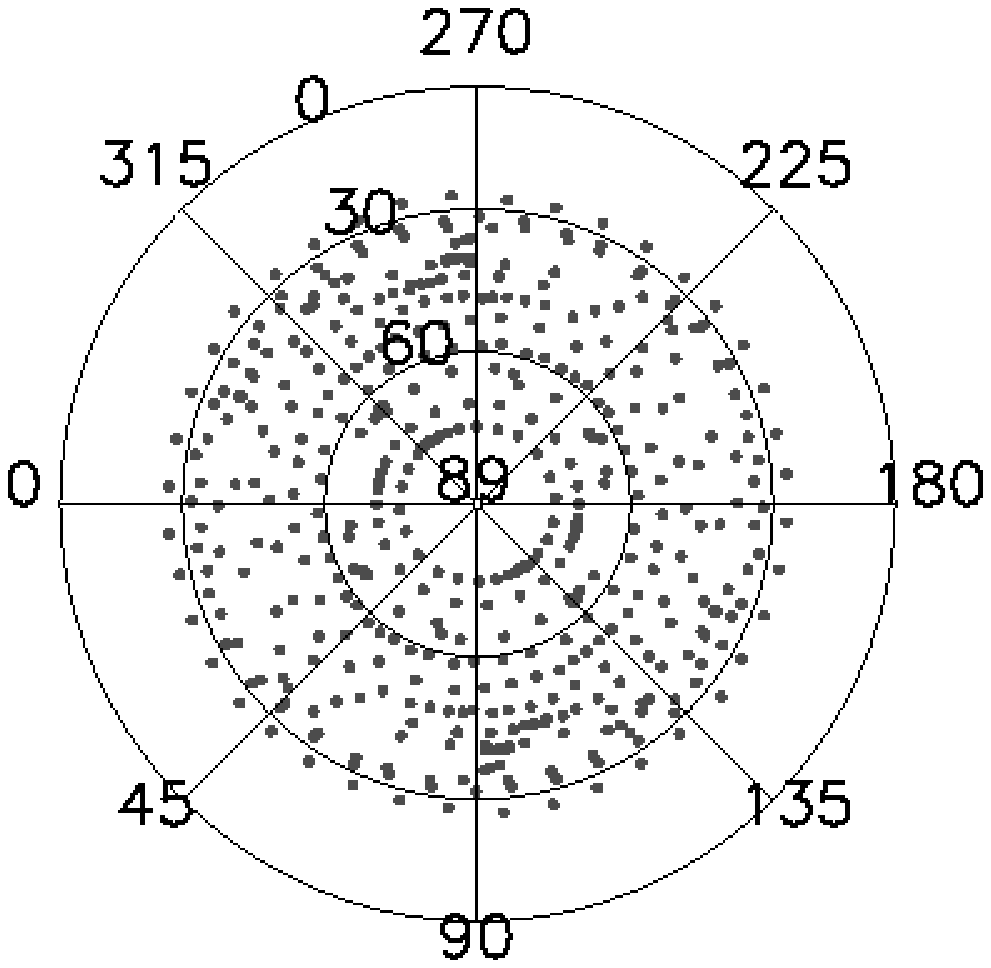}
\caption{
Polar projections showing the overshoot region with the uniformly
distributed positions $(\lambda_\mathrm{s},\phi_\mathrm{s})$ of initial
perturbations (\emph{left}) and a surface distribution with the
positions $(\lambda_\mathrm{e},\phi_\mathrm{e})$ of erupting flux tubes
resulting from their non-linear evolution in the convection zone
(\emph{right}).
In this example with $B_\mathrm{s}= 1.2\cdot10^5\,\mathrm{G}$ preferred
longitudes are discernable at intermediate latitudes
$\lambda_\mathrm{e}\sim 45\degr$ near the quadrature longitudes; a
number of flux tubes do not emerge due to their stable initial
configuration.
The star rotates in counter-clockwise direction and the companion star 
is located in the direction $\phi= 0$.
}
\label{tubedist.pic}
\end{figure}

The linear stability analysis in HS{\it I} describes the response of a
toroidal equilibrium flux tube to small perturbations inside the
overshoot region and its evolution prior to its penetration to the
convection zone above.
In the evolution of rapidly rising flux loops in the superadiabatical
region non-linear effects of the tube dynamics become important, which 
have to be followed by numerical simulations.
According to the results in HS{\it I}, we consider flux tubes with
initial magnetic field strengths between $7\cdot10^{4}\,\mathrm{G}\le
B_\mathrm{s}\le 2\cdot10^{5}\,\mathrm{G}$.
The initial flux rings are located in the middle of the overshoot
region at latitudes between $0\degr\le \lambda_\mathrm{s}\le 80\degr$
parallel to the equatorial plane and liable to an undulatory
(Parker-type) instability \citep[e.g.,][]{1995GAFD...81..233}.
The simulations start with a perturbation of a flux ring and cover up 
to $32\,\mathrm{yrs}$ of its evolution in the convection zone.
In order to sample the influence of tidal effects on rising flux tubes,
we carry out a large number of simulations in which each tube is
subject to a perturbation \emph{localised around a different
longitude}, $\phi_\mathrm{s}$.
This is realised by an in-phase superposition of sinusoidal
displacements, $\sin m\phi$, with azimuthal wave numbers $m= 1\ldots 5$
and amplitudes of one percent of the local pressure scale height
(Fig.~\ref{excite.pic}).
For simulations pertaining to a given
$(B_\mathrm{s},\lambda_\mathrm{s})$ configuration, this specific
perturbation is successively shifted in longitude; the interval $0\le
\phi_\mathrm{s}< 180\degr$ is sampled with 20 simulations, each
separated by $\Delta\lambda_\mathrm{s}= 9\degr$.  Owing to the
azimuthal $\pi$-periodicity of the underlying problem, the resulting
pattern of flux tube emergence is then extended to the interval
$180\degr\le \phi_\mathrm{s} < 360\degr$.
The regularly distributed positions
$(\lambda_\mathrm{s},\phi_\mathrm{s})$ of the localised perturbations
are shown as points in the polar projection of the overshoot region in
the left panel of Fig.~\ref{tubedist.pic}.
The evolution of rising flux tubes is followed close to their eruption
at the stellar surface, resulting in spot distributions
$(\lambda_\mathrm{e},\phi_\mathrm{e})$ like the example shown in the
right panel of Fig.~\ref{tubedist.pic}.
A complete set of simulations represents a non-linear mapping from the
bottom of the convection zone to the top, assuming uniform
distributions for the perturbations over longitude and for the flux
rings over latitude in the overshoot region.
The simulations stop at about $0.98\,R_{\sun}$, where the `thin flux 
tube approximation' ceases to be valid.
Since the rise of the tube's crest through the uppermost parts of the
convection zone is very fast and nearly radial, there is hardly a
difference between its final position in the simulation and the
position of eruption at the stellar surface.

A toroidal isentropic flux tube embedded in the tidally distorted 
stratification of a binary star actually represents a non-equilibrium 
configuration, since it exhibits a density contrast along the tube axis
which entails a small deviation from mechanical equilibrium.
Following the linear stability analysis in HS{\it I}, the dynamical
tube evolution due to this azimuthal density variation proceeds on very
long time scales of several thousand days, which are not relevant for
the eruption of magnetic flux at the stellar surface..
Since the perturbation of the flux ring due to the density variation
along the tube is much smaller than the perturbation arising from the
superimposed displacement (described above) and furthermore identical
for all simulations at a given latitude, its effect, if any, on the
longitudinal distribution of erupting flux tubes is expected to be
neglectable.

\section{Results}
\label{resu}

\subsection{Eruption times}
\label{erti}
\begin{figure}
\includegraphics[width=\hsize]{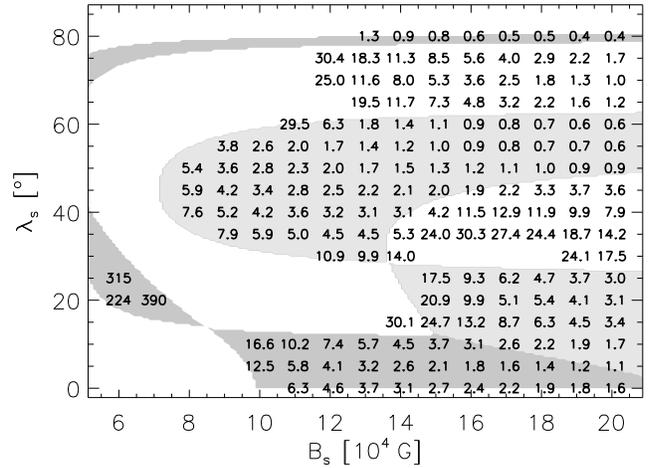}
\caption{
Eruption times $t_\mathrm{e}$ in years, where the position
$(B_\mathrm{s},\lambda_\mathrm{s})$ in the diagram refers to the
initial flux tube configuration in the overshoot region.
According to the linear stability analysis in HS{\it I}, grey shaded
regions mark domains of Parker-type instabilities with dominating
azimuthal wave numbers $m= 1$ (\emph{dark}) and $m= 2$ (\emph{light}).
Configurations in white regions are linearly stable, but yield
eruptions due to the non-linear flow instability.
} 
\label{timedist.pic}
\end{figure}

The eruption times, $t_\mathrm{e}$, defined as the elapsed time between
the initial perturbation of a flux tube and its emergence at the
stellar surface, are found to be of the order of several months to
years, in a few cases (near the instability boundaries) up to decades.
Their dependence on the initial $(B_\mathrm{s},\lambda_\mathrm{s})$ 
configurations (Fig.~\ref{timedist.pic}) is roughly proportional to the
respective growth times of the linear instability in the overshoot 
region, which are, according to the analysis in HS{\it I}, dominated by 
Parker-type instabilities with azimuthal wave numbers $m= 1$ and $2$.
On the other hand, flux tubes in linearly stable initial
configurations, e.g., at $\lambda_\mathrm{s}\simeq 65\ldots75\degr$ or
in the stability island at intermediate latitudes and high field
strengths, are found to become unstable by another mechanism, which we
ascribe to the interaction of a flux tubes with the environment due to
its hydrodynamic drag once the relative tangential flow velocity
exceeds the phase velocity of a perturbation; a detailed description of
this flow instability will be given in a subsequent paper.
This second-order effect is not included in the framework of the linear
stability analysis in HS{\it I}, but nevertheless appears in the course
of non-linear simulations and leads to the formation of rising flux
tubes on time scales comparable to those of the Parker-type
instability.
At high magnetic field strengths this instability mechanism contributes
a substantial fraction to the total number of erupting flux tubes.
In contrast, simulations with initial configurations in the instability
islands at low field strengths and $\lambda_\mathrm{s}\sim 20\degr$
result in eruption times of several hundred years and thus are probably
irrelevant.

The simulated rise of flux tubes through the convection zone does not
always proceed in a continuous manner.
Some flux tubes become unstable and form rising loops in the overshoot 
region, but then pass through a dynamical stage of quasi-stability 
where their radial rise temporarily comes to a halt before they resume 
after some several hundred days their ascent through the convection 
zone toward the surface.

\subsection{Latitudinal distributions}
\label{ladi}
\begin{figure}
\includegraphics[width=\hsize]{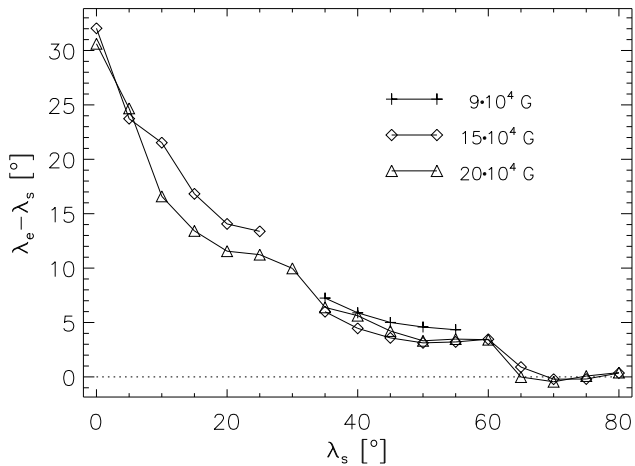}
\caption{
Poleward deflection, $\Delta\lambda= \lambda_\mathrm{e} -
\lambda_\mathrm{s}$, of erupting flux tubes as a function of their
initial latitude $\lambda_\mathrm{s}$ in the overshoot region.
Symbols mark different initial field strengths, $B_\mathrm{s}$.
} 
\label{latdist.pic}
\end{figure}

Whereas sunspots exclusively appear in an equatorial belt between about
$\pm 35\degr$ latitude, fast-rotating stars frequently show spots also
at intermediate, high, and even polar latitudes.
In the framework of the magnetic flux tube model, this behaviour is
explained by the influence of the Coriolis force, which acts on the 
internal gas flow along the tube axis and causes a poleward deflection
of the tube \citep{1992A&A...264L..13S}.
Figure \ref{latdist.pic} shows the latitude excursion, $\Delta\lambda=
\lambda_\mathrm{e} - \lambda_\mathrm{s}$, i.e., the difference between
the latitude of emergence, $\lambda_\mathrm{e}$, and the starting
latitude in the overshoot region, for flux tubes with different initial
magnetic field strengths, $B_\mathrm{s}$.
The deflection is largest for flux tubes starting at low latitudes.
It decreases considerably with increasing $\lambda_\mathrm{s}$ and 
eventually drops to zero at high latitudes.
Flux tubes starting above $\lambda_\mathrm{s}\gtrsim 60\degr$
essentially do not show any deflection.
The poleward deflection depends only marginally on the initial magnetic
field strength.
At moderate latitudes, $\Delta \lambda$ becomes somewhat smaller for
increasing $B_\mathrm{s}$, because the larger buoyancy force leads to a
more radial rise through the convection zone.

\subsection{Longitudinal distributions}
\label{lodi}
\begin{figure}
\includegraphics[width=\hsize]{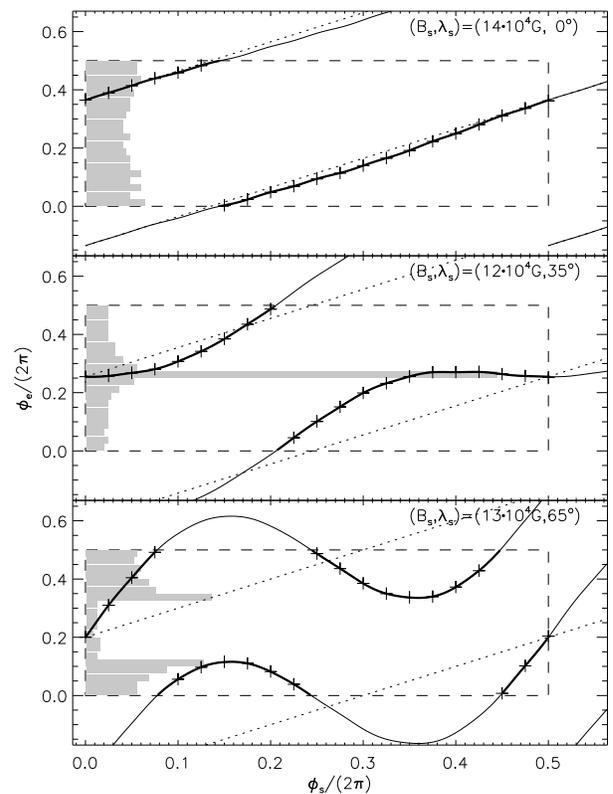}
\caption{
Distributions of eruption longitudes, $\phi_\mathrm{e}$, as a function
of the initial perturbation longitude, $\phi_\mathrm{s}$.
Crosses mark the results of simulations, solid lines the (interpolated)
non-linear functions $\phi_\mathrm{e}(\phi_\mathrm{s})$ in the binary
problem, and dotted lines the linear functions in the corresponding
single star problem.
The grey histograms at the left side show the resulting distributions 
of erupting flux tubes binned to $9\degr$-wide longitudinal intervals 
(abscissa gives normalised values).
The distributions show (from \emph{top} to \emph{bottom}) a rather flat
distribution, a clearly preferred longitude, and a favoured longitude
range for emergence, respectively.
} 
\label{sire_przp.pic}
\end{figure}

Figure \ref{sire_przp.pic} shows examples of longitudinal distributions
of erupting flux tubes in a binary star.
Owing to the wave-like character of the instability, the longitude of 
emergence, $\phi_\mathrm{e}$, differs from the formation longitude of a 
rising loop in the overshoot region.
In a single star, an azimuthal shift of the localised perturbation by
an offset $\Delta \phi_\mathrm{s}$ yields the same offset in the
longitude of emergence.
Under the action of tidal forces in a deformed binary star, however,
the dynamics and evolution of a flux tube depend on the value of
$\phi_\mathrm{s}$.
The longitudes of emergence thus exhibit a deviation from the uniform
distribution for a single star, with the amplitude of the
$\pi$-periodic deviation depending on the
$(B_\mathrm{s},\lambda_\mathrm{s})$ values of the initial flux ring.
This results in a non-uniform longitudinal distribution of erupting
flux tubes (represented in Fig.~\ref{sire_przp.pic} by the histograms
at the left side of each panel), which show clear indications for
preferred longitude ranges of flux eruption: beside cases with only
marginal, if any, longitudinal non-uniformity, there are cases of
highly peaked distributions as well as distributions where a broad
longitudinal interval is preferred.

\subsubsection{High latitudes ($\lambda_\mathrm{s}= 65\ldots80\degr$)}
\label{hila}
\begin{figure}
\includegraphics[width=\hsize]{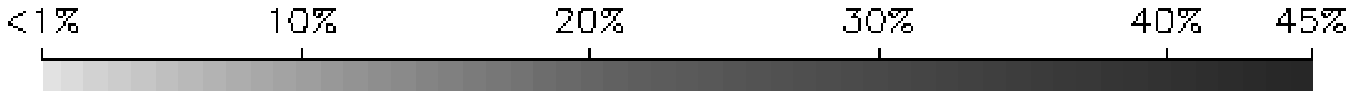}
\includegraphics[width=\hsize]{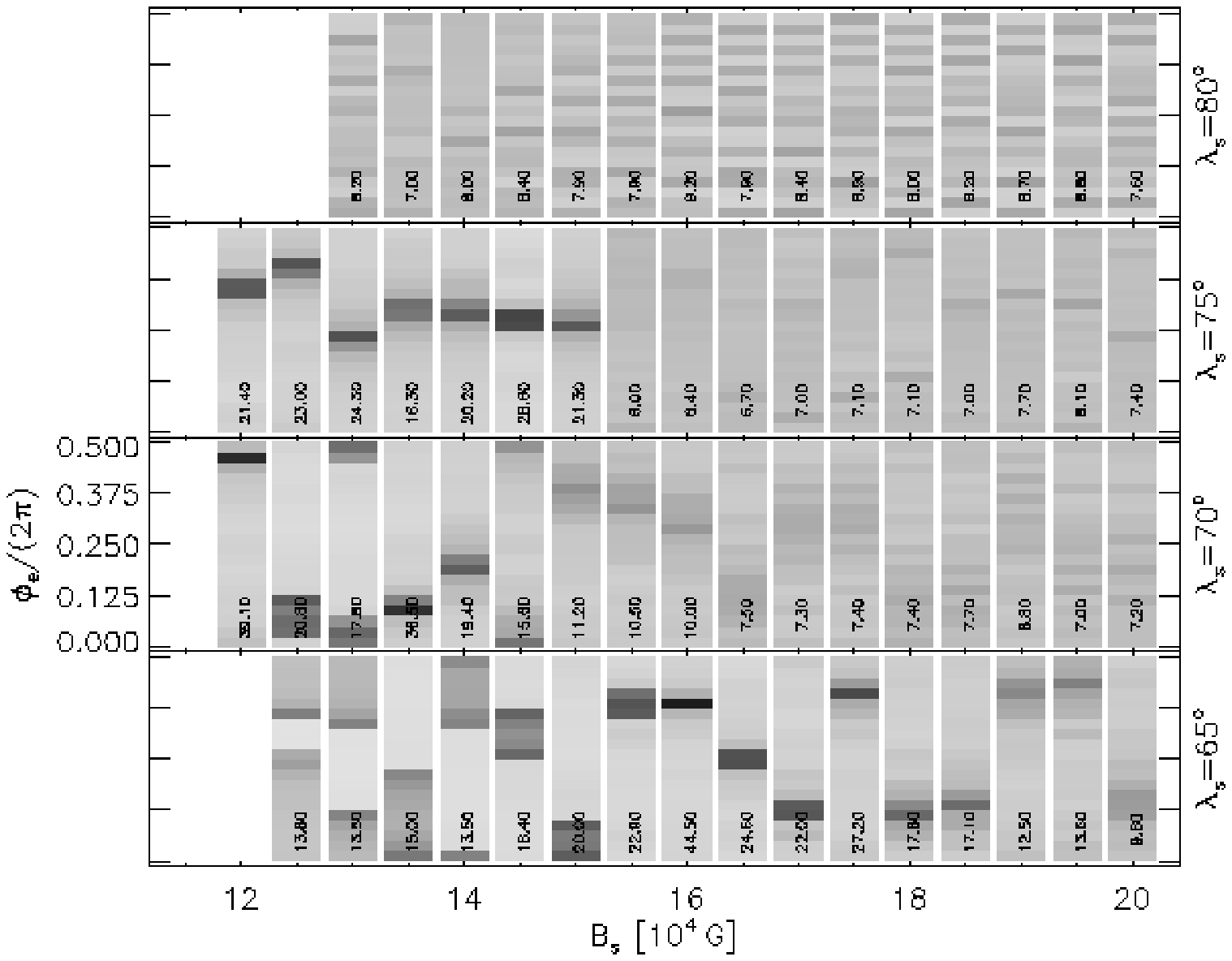}
\caption{
Longitudinal distributions of erupting flux tubes with high initial
latitudes, $\lambda_\mathrm{s}= 65\ldots80\degr$. 
The shading indicates the amount of clustering inside $9\degr$-wide
longitudinal intervals normalised to the total number of flux tubes for
each parameter pair $(B_\mathrm{s},\lambda_\mathrm{s})$; the numbers
give the maximum values for each case.
Because of the $\pi$-periodicity in longitude, only the range $0\le 
\phi_\mathrm{e}/(2\pi)< .5$ is shown.
White regions indicate stable, non-erupting flux tubes.
} 
\label{londisthl.pic}
\end{figure}

Figure \ref{londisthl.pic} shows the longitudinal patterns of emergence
for high initial latitudes, based on the corresponding histogram for
each $(B_\mathrm{s},\lambda_\mathrm{s})$ configuration.
The shading indicates the amount of longitudinal clustering of erupting 
flux tubes.
For $\lambda_\mathrm{s}\approx 80\degr$ we have Parker-type
instabilities with the azimuthal wave number $m= 1$ and almost flat
distributions in longitude.
For latitudes between $\lambda_\mathrm{s}\approx 65\ldots75\degr$ and
moderate field strengths considerable asymmetries appear in the form of
highly peaked distributions or broad preferred intervals.
In some cases, up to $45\%$ of all flux tubes of a
$(B_\mathrm{s},\lambda_\mathrm{s})$ configuration emerge within a
longitude interval of only $9\degr$ width.
Apart from a decreasing tendency for clustering with increasing field
strength, there is no concise systematic dependency of the longitude
distribution of these features on the initial
$(B_\mathrm{s},\lambda_\mathrm{s})$ parameters.

\subsubsection{Intermediate latitudes ($\lambda_\mathrm{s}=
30\ldots60\degr$)}
\label{inla}
\begin{figure}
\includegraphics[width=\hsize]{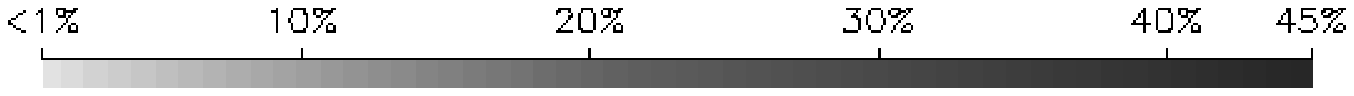}
\includegraphics[width=\hsize]{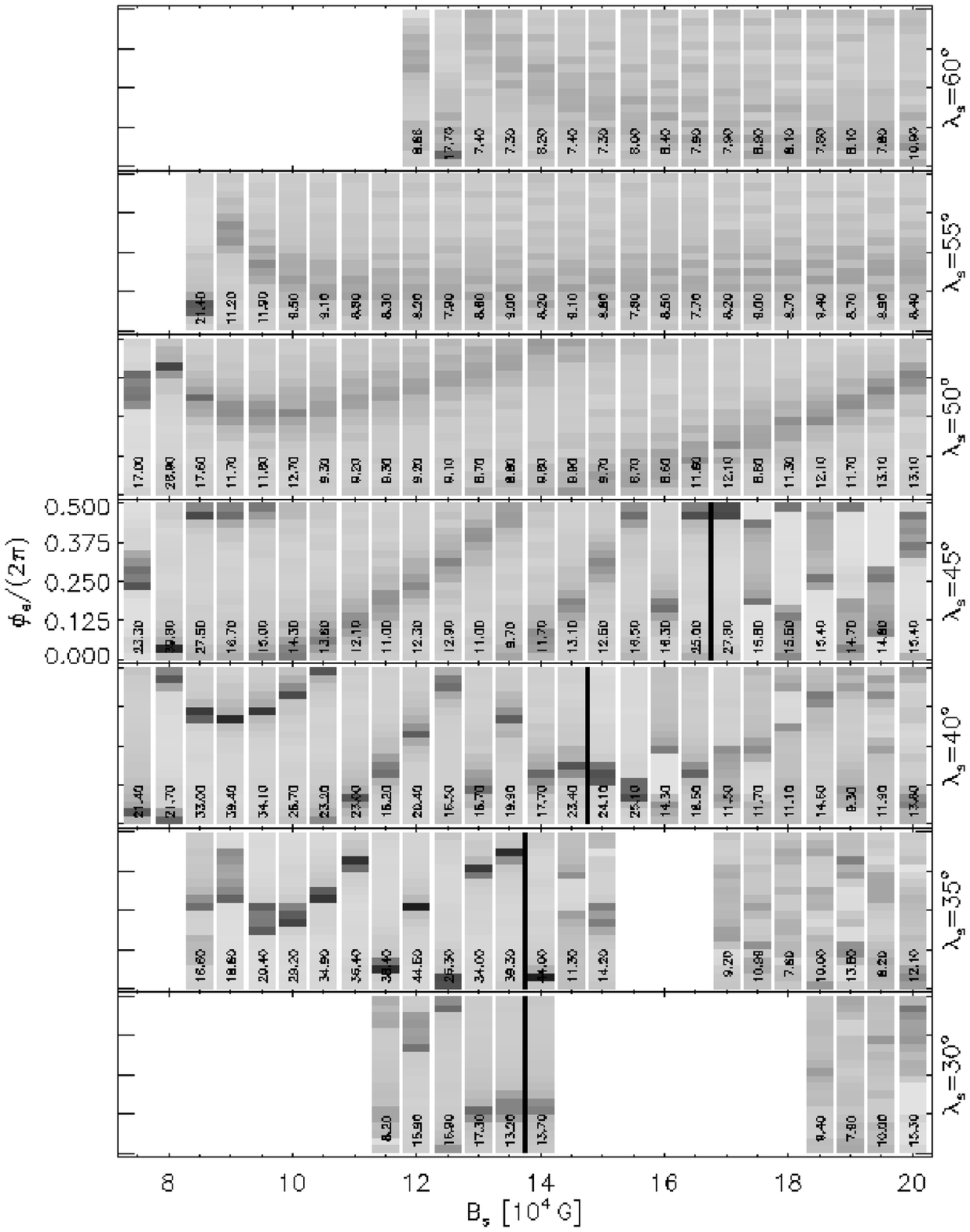}
\caption{
Longitudinal distributions of erupting flux tubes with intermediate
initial latitudes $\lambda_\mathrm{s}= 30\ldots60\degr$.
The vertical black lines divide the domains of linear instabilities
according to HS{\it I}: the upper left domain is dominated by
Parker-type instabilities with $m= 2$, whereas the region to the right
shows the flow instability.
White regions indicate stable, non-erupting flux tubes.
} 
\label{londistil.pic}
\end{figure}

The evolution of flux tubes starting at intermediate latitudes are
significantly affected by tidal effects.
Figure \ref{londistil.pic} shows longitudinal distributions of erupting
flux tubes starting between $\lambda_\mathrm{s}= 30\ldots60\degr$ in
the overshoot region.
In the domain of Parker-type instabilities with the azimuthal wave 
number $m= 2$ (left to the vertical black lines), the concentration and 
orientation of clusters of emerging flux tubes depends systematically 
on the initial values $(B_\mathrm{s},\lambda_\mathrm{s})$.
The most prominent clusters occur for $\lambda_\mathrm{s}\approx
35\ldots40\degr$ and $B_\mathrm{s}\approx
8\ldots14\cdot10^4\,\mathrm{G}$.
An important aspect is the orientation of preferred longitudes with 
respect to the direction of the companion star at $\phi= 0$.
Figure \ref{londistil.pic} shows that clusters basically show up at all 
orientations depending on the initial configuration.
Since there is no unique orientation for all configurations we use the 
term \emph{preferred longitude} in relation to the clustering of 
erupting tubes for an individual $(B_\mathrm{s},\lambda_\mathrm{s})$ 
configuration.
For increasing magnetic field strength near the instability threshold,
the clusters are shifted toward smaller longitudes (retrograde with
respect to the rotation of the system), whereas for high $B_\mathrm{s}$
this trend reverses and the clusters shift toward larger longitudes
(prograde).
The slope of this longitudinal shift as a function of field strength 
depends on the starting latitude of the tube.
At small $\lambda_\mathrm{s}$, the slope is rather large and an
increase of $B_\mathrm{s}$ by about $5\cdot10^3\,\mathrm{G}$ can result
in an azimuthal shift of the cluster by $90\degr$.
The slope decreases considerably with increasing starting latitude and 
approaches zero at $\lambda_\mathrm{s}\approx 55\degr$.
Here the orientation of clusters is apparently independent of the 
magnetic field strength, thus representing a kind of \emph{fixed}
preferred longitudes. 
However, in this case the clustering is rather weak.
The shift rate seems to be related to the amount of clustering, with 
more concentrated clusters showing larger shift rates.

\subsubsection{Low latitudes ($\lambda_\mathrm{s}= 0\ldots25\degr$)}
\label{lola}
Longitudinal distributions of erupting flux tubes starting at
equatorial latitudes ($\lambda_\mathrm{s}\le 10\degr$) and liable to
the Parker-type instability with wave number $m= 1$ do not show
significant non-uniformities.
For $\lambda_\mathrm{s}= 15\ldots25\degr$, flux tubes become unstable
above $B_\mathrm{s}\simeq 1.4\cdot10^5\,\mathrm{G}$ with the dominating
wave mode $m= 2$.
This domain of initial configurations is characterised by a particular
evolutionary behaviour: flux tubes with the same
$(B_\mathrm{s},\lambda_\mathrm{s})$ do not exhibit a systematic
continuous dependence of the emergence longitude on $\phi_\mathrm{s}$
like in Fig.~\ref{sire_przp.pic}, but an irregular distribution.
This aspect of the simulation results is discussed in more detail in 
Sect.~\ref{disc}.

\subsection{Surface distributions}
\label{sudi}
\begin{figure}
\includegraphics[width=\hsize]{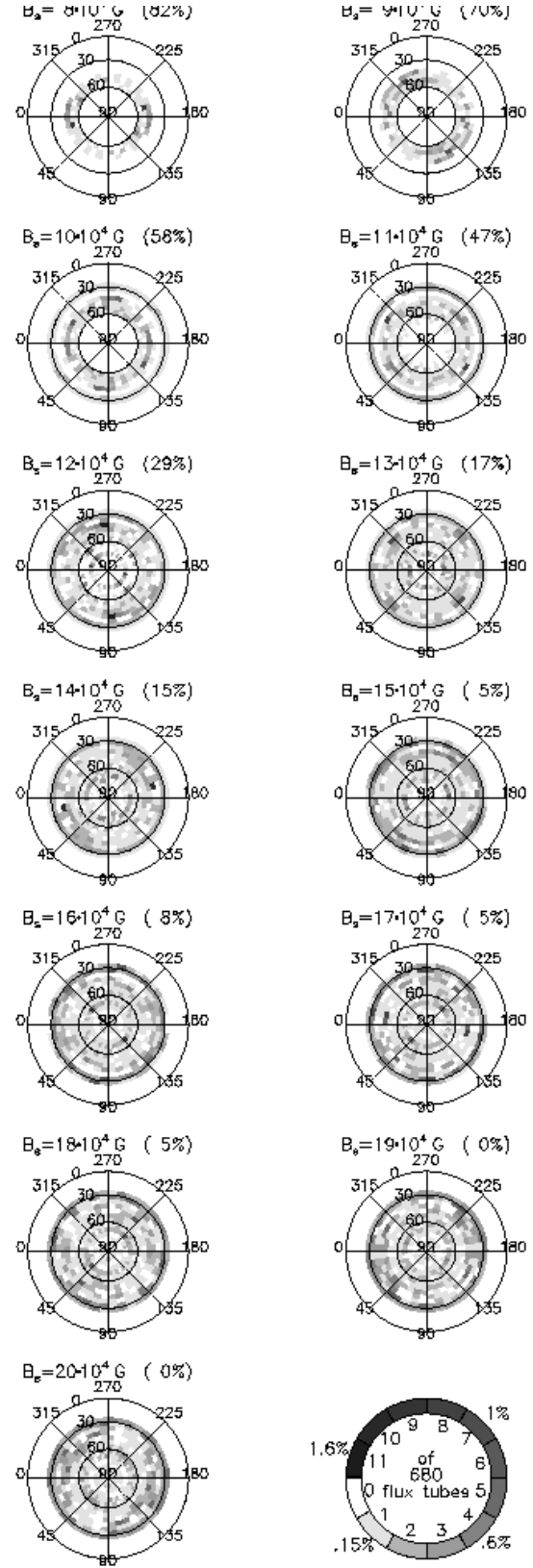}
\caption{
Surface distributions of erupting flux tubes for different initial 
field strengths, $B_\mathrm{s}$.
The star rotates in counter-clockwise direction, $\phi= 0$ is the 
direction toward the companion star.
The shading indicates the number of flux tubes erupting in a
$(\Delta\phi_\mathrm{e},\Delta\lambda_\mathrm{e})=
(9\degr,5\degr)$-wide surface area.
The values in brackets give the relative number of flux tubes which do 
not erupt at the surface due to their stable equilibrium in the 
overshoot region.
}
\label{surfdist.pic}
\end{figure}

Figure \ref{surfdist.pic} shows polar projections of the surface
distributions of erupting flux tubes with different initial field
strengths, $B_\mathrm{s}$, combining both longitudinal and latitudinal
information with the grid of simulations $(\Delta\phi_\mathrm{s},
\Delta\lambda_\mathrm{s})= (9\degr,5\degr)$ shown in the left panel of
Fig.~\ref{tubedist.pic}.
Owing to the poleward deflection and the azimuthal non-uniformities
induced by tidal effects, the surface distributions deviate
considerably form the uniform distribution of initial flux tubes at the
bottom of the convection zone and typically show features on opposite
sides of the stellar hemisphere resulting from the dominating
$\pi$-periodicity of tidal effects.
While some parts of the surface are avoided by erupting flux tubes, 
like the equatorial belt below about $25\degr$, they gather in other
regions to form clusters of flux eruption.
For low magnetic field strengths, only a few flux tubes starting from
intermediate latitudes reach the surface, whereas the majority of tubes
at other latitudes does not leave the overshoot region owing to their
stable initial configuration.
For increasing field strengths more flux tubes become unstable and 
contribute to the overall surface pattern erupting particularly at 
moderate and high latitudes.
The flux eruption at high latitudes is mainly due to the flow
instability, which results in clusters, if any, covering broad
longitudinal intervals.
Flux tubes erupting at moderate latitudes are due to Parker-type
instabilities with wave number $m= 1$ starting at low initial
latitudes.
These configurations hardly show longitudinal asymmetries.
However, flux tubes starting at somewhat higher latitudes with $m= 2$
experience a smaller poleward deflection and emerge at moderate
latitudes, too, showing considerably non-uniform longitudinal
distributions.
At intermediate latitudes and high field strengths the extended and 
less pronounced clusters are due to the flow instability.
The systematic dependence of preferred longitudes, which is discernible
in the distributions shown in Fig.~\ref{londistil.pic}, is less evident
in the surface patterns, because, owing to the variable poleward
deflection, flux tubes from different initial latitudes emerge in the
same surface area.
Since the orientation of clusters in distributions with the same field
strength $B_\mathrm{s}$ but different latitudes $\lambda_\mathrm{s}$
are usually not the same, their superposition in the surface pattern
sometimes leads to broader emergence patterns.
However, particularly at low field strengths, the existence of
considerable clusters at preferred longitudes remain, while for higher
field strengths significant non-uniformities are still discernible in
the overall surface patterns.

\subsection{Dependence on system period}
\label{dosp}
\begin{figure}
\includegraphics[width=\hsize]{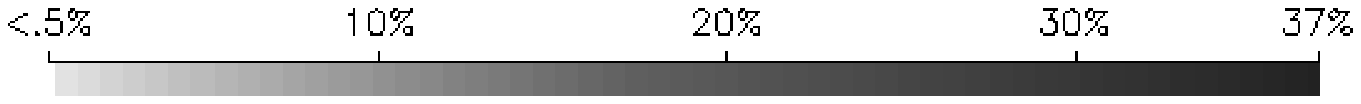}
\includegraphics[width=\hsize]{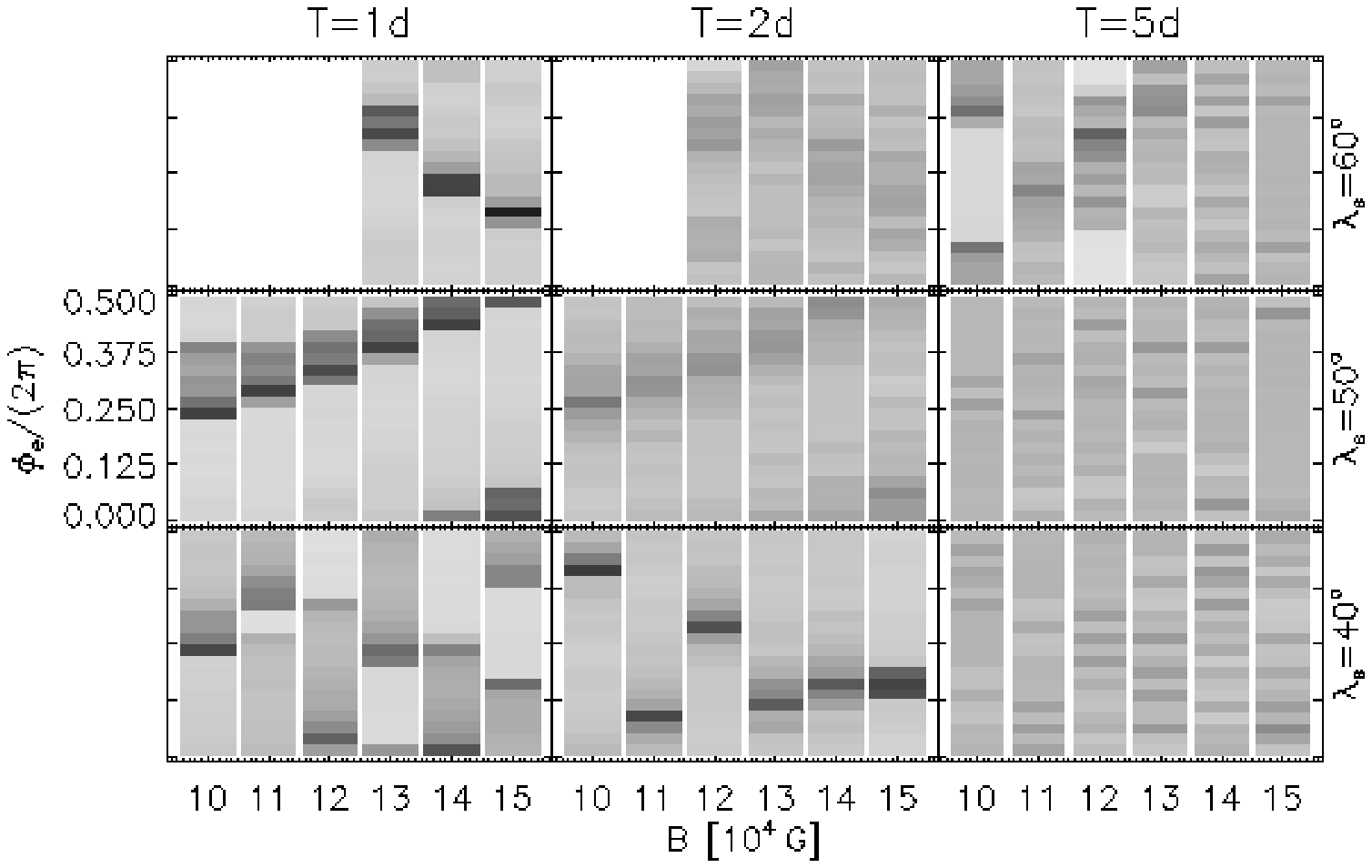}
\caption{
Longitudinal distributions of erupting flux tube for systems with
orbital periods $T= 1, 2$ and $5\,\mathrm{d}$ (from \emph{left} to
\emph{right}).
Shown are distributions for flux tubes starting at intermediate
latitudes.
} 
\label{loncomp.pic}
\end{figure}

We have investigated the influence of the magnitude of the tidal
effects by considering systems with orbital periods $T= 1\,\mathrm{d}$
and $5\,\mathrm{d}$, which correspond to binary separations $a=
5.3\,R_{\sun}$ and $15.5\,R_{\sun}$, respectively.
Figure \ref{loncomp.pic} shows a comparison of longitudinal
distributions of eruption for flux tubes with intermediate initial
latitudes and field strengths $B_\mathrm{s}=
1\ldots1.5\cdot10^5\,\mathrm{G}$.
In the case $T= 5\,\mathrm{d}$, most longitudinal distributions show
hardly any significant clusters of eruption.
Merely for $\lambda_\mathrm{s}= 60\degr$ there are some
non-uniformities, which indicate that tidal effects can still be
effective for some initial tube configurations.
The case $T= 2\,\mathrm{d}$, the reference case discussed in detail
above, shows the existence of considerable emergence clusters and the
systematic dependence of their orientation on the initial tube
configuration.
For the shorter orbital period $T= 1\,\mathrm{d}$ the degree of
clustering increases.
However, for some initial tube configurations there is a transition
from highly peaked distributions toward more extended preferred
intervals.
In contrast to highly peaked clusters these patterns of eruption are 
rather characterised by an avoidance of certain longitudinal intervals.

\section{Discussion}
\label{disc}
Our results show that the presence of the companion star in a close
binary system can considerably alter the evolution and surface
distribution of erupting magnetic flux tubes.
The actual influence of tidal effects on a rising flux tube depends on 
its initial magnetic field strength, $B_\mathrm{s}$, and latitude,
$\lambda_\mathrm{s}$. 
This includes the stability properties of the initial flux ring in the
overshoot region, i.e., the instability mechanism and the azimuthal 
wave number, $m$, of the dominating unstable eigenmode, which governs 
the character of the subsequent non-linear evolution.
In our range of parameters, only single-loop tubes (with $m= 1$) and 
double-loop tubes (with $m= 2$) appear.

The properties of erupting flux tubes can roughly be classified
according to their linear instability mechanism as determined in HS{\it
I}.
In the domain of Parker-type instabilities, erupting single-loop tubes 
starting at either low or very high latitudes tend to form uniform 
longitudinal patterns of eruption, whereas double-loop tubes result in 
clusters of eruption concentrated in narrow longitudinal intervals.
This specific behaviour is presumably due to the congruency between the
double-loop geometry of the rising tube and the $\pi$-periodicity of 
the tidal effects, which enables a \emph{resonant} interaction.
Although there is no globally unique orientation for the clusters of
eruption, the longitudinal shift of the clusters (usually prograde with
increasing $B_\mathrm{s}$) can become very small and even vanish
leading to \emph{fixed} preferred longitudes at a certain latitude.
Tubes erupting due to the flow instability show a strong susceptibility
to tidal effects leading to extended preferred intervals.
In contrast to the domain of the Parker-type instability with $m= 2$, 
the orientations of the preferred intervals in this case lack a 
systematic dependence on initial tube parameters.
Particularly at high latitudes and high field strengths the flow 
instability is responsible for the majority of erupting flux tubes.
The dependence of the evolution of flux tubes on the instability
mechanism (with similar initial configurations otherwise) indicates
that the formation of non-uniformities at the surface is not only a
matter of the tidal forces acting during the non-linear rise of the
flux tubes through the convection zone, but also affected by stability
properties in the overshoot region.
While the Parker-type instability is basically a buoyancy-driven
mechanism and thus susceptible to perturbations of gravity and stellar
structure, the flow instability is more related to the tube parameters,
namely its diameter and the relative velocity perpendicular to the
environment.

For some $(B_\mathrm{s},\lambda_\mathrm{s})$ combinations the set of
erupting flux tubes does not yield a continuous function
$\phi_\mathrm{e} (\phi_\mathrm{s})$ like the ones shown in
Fig.~\ref{sire_przp.pic}.
Instead, they yield either piecewise steady functions, which frequently
alternate between two branches or irregular distributions, which show
no obvious relation to the original shift of perturbation longitudes,
$\phi_\mathrm{s}$.
These particular configurations with irregular distributions are
limited to $B_\mathrm{s}\gtrsim 1.5\cdot10^5\,\mathrm{G},
\lambda_\mathrm{s}\simeq 15\ldots35\degr$, a parameter domain mainly
showing a Parker-type instability with $m= 2$, which roughly coincides
with the region of retarded flux eruption described in
Sect.~\ref{erti}.
Because of the dominating unstable wave mode with $m= 2$, two loops of
similar size form at opposite sides of the initially toroidal flux tube
in the overshoot region and enter the convection zone, whereby the
wave-like character of the instability causes a longitudinal
propagation.
Non-linearities lead to slightly different evolutions of the two loops,
so that the early stage of the development is characterised by
consecutive changes of the dominance between both loops until one of
them prevails and eventually emerges.
In this stage, the time delays of rising flux loops described above
entail longitudinal offsets owing to their wave-like propagation.  Flux
tubes starting with different values $\Delta\phi_\mathrm{s}$ may suffer
different time delays before they resume their rise and emerge at the
surface.
The originally continuous phase shift $\Delta\phi_\mathrm{s}$ imprinted
on a set of flux tubes is thus annihilated and the resulting
longitudinal distribution irregular.  Since for any
$(B_\mathrm{s},\lambda_\mathrm{s})$ configuration only twenty
simulations (per semi-hemisphere) are accomplished, the irregular
distributions can however exhibit non-uniformities due to the small
statistical sample.
For a large number of simulations we would expect a rather uniform
distribution.

The trajectories of rising flux tubes exhibit the poleward deflection
typical for fast rotators, which is caused by the dominance of the 
Coriolis force over the buoyancy force \citep{1992A&A...264L..13S}.
The latitudinal deflection leads to an overlap of patterns of erupting
flux tubes with different initial latitudes, which results in a
superposition of clusters at different longitudes thus broadening the
range of preferred longitudes in the surface distributions.
The deflection is largest for flux tubes starting at equatorial and low
latitudes and decreases for intermediate and high initial latitudes.
An explanation for the observed high-latitude and polar spots in the
framework of the present erupting flux tube model thus requires large
amounts of magnetic flux at high latitudes in the overshoot region,
either generated by a suitable dynamo mechanism or transported by
large-scale motions inside the convection zone.
Alternatively, flux erupting at lower latitudes may be transported 
toward the pole by meridional motions \citep{2001ApJ...551.1099S}.

The investigation of binary systems with different orbital periods,
i.e., distance between the components, shows that the non-uniformities
become more pronounced with stronger tidal effects.
The orbital period of about $T\sim 5\,\mathrm{d}$ for the onset of
clustering is in agreement with the results from
\citet{1995AJ....109.2169H}.
In general, the non-uniformities increase for smaller orbital periods,
but this trend is accompanied by a transition from clusters localised 
in narrow longitudinal intervals to broader intervals, up to major 
fractions of the stellar circumference.
The number of cases with irregular longitudinal distributions increases
for smaller orbital periods.
This is in accordance with our interpretation that, with a closer
proximity of the companion star, the resonant interaction between the
tube and the tidal effects becomes more efficient and eventually ends
up in an analogy to the `resonant catastrophe' of a driven oscillator.

The action of tidal effects on the dynamics of magnetic flux tubes is
basically a \emph{cumulative} effect, which proceeds and adds up during
the entire rise from the overshoot region up to the surface.
During this time, an individual flux tube is subject to further
interactions with, e.g., turbulent convective motions, differential
rotation, meridional flows, and with other concentrations of magnetic
flux.
Although these additional interactions may modify the evolution and
alter the final location of eruption to some extent, we do not expect
them to severely diminish the characteristic \emph{non}-axisymmetric
influence of the companion star since they are presumably axisymmetric
on average.
However, it should be kept in mind that the applied thin flux tube
model represents a rough simplification of the magnetic field 
structure in convective stellar envelopes.

The distributions of flux eruption shown above allow \emph{statistical}
predictions, which apply to large numbers of erupting flux tubes with 
similar initial configurations.
Our simulations are based on flux tubes evenly distributed in latitude 
as well as on perturbations evenly distributed in longitude.
A changing situation in the overshoot region, e.g., a dependence of the
amount of magnetic flux on latitude due to a particular dynamo
mechanism or a dependence of tube perturbations on longitude due to a
dependence of convective motions on tidal effects, could, in principle,
be taken into account by folding the specific initial condition with
the present non-linear mapping of rising flux tubes from the bottom of
the convection zone to the top.
Although the frequently observed occurrence of spot clusters at
opposite sides of stellar hemispheres can consistently be explained by
the dominance of the $\pi$-periodic contributions of tidal effects, our
time-independent patterns of emergence do not immediately allow a
description of the frequently observed azimuthal migration of spot
clusters in time \citep{1994A&A...281..395S, 1995A&A...301...75R,
2000A&A...358..624R, 1998A&A...336L..25B}.
However, a non-stationary dynamo mechanism with latitudinal propagation
of a dynamo wave would lead to a time- and latitude-dependent field
strength and magnetic flux, so that the systematic dependence of the
cluster longitudes on field strength shown in Fig.~\ref{londistil.pic}
could transform into a longitudinal migration of the spot clusters in
time.

We considered a binary system with two solar-type main sequence 
components to investigate the basic effects of tidal interaction on the
evolution of erupting flux tubes.
Although the results shown here are consistent with some observations,
there are observed spot properties which are not explained by our
results, like spots at very low latitudes or fixed spot clusters around
the substellar point \citep{2001A&A...376.1011L, 2002A&A...386..583L,
2002A&A...389..202O}.
RS CVn systems, however, typically consist of at least one (sub-)giant
component with a much deeper convective envelope than the Sun, leading
to the assumption that the formation and orientation of spot clusters
are not exclusively dependent on the binary separation alone, but also
in a more complicated way on the type and structure of the star.
The deep convection zone of a giant, for example, will probably modify
the properties of patterns of eruption owing to the cumulative
character of the interaction between the rising flux tube and the
companion star.

\section{Conclusion}
\label{conc}
Erupting flux tubes in binary stars are considerably affected by the
tidal influence of the companion star.
Although the magnitude of tidal effects -- the tidal force and the
deviation of the stellar structure from spherical symmetry -- are
rather small for the systems considered here, they are capable to alter
the surface distributions of erupting flux tubes.
This results in clusters of emerging flux tubes at preferred longitudes
and in the formation of spot clusters on opposite sides of the active 
component.
We consider two main reasons for this behaviour:
\begin{enumerate}
\item
Contrary to single-loop tubes, which are not strongly affected, the
geometry of double-loop tubes (with azimuthal wave number $m= 2$)
allows of an efficient \emph{resonant interaction} with the congruent
$\pi$-periodic structure of the dominating tidal effects.
\item
Since the rise of magnetic flux tubes from the bottom of the convection 
zone until eruption at the surface takes about several months or years,
the \emph{cumulation of the tidal effects} leads to a strong influence 
on the resulting emergence pattern.
\end{enumerate}

The distributions of erupting flux tubes shown here are based on
uniformly distributed initial conditions (distributions of flux tubes
and their localised perturbation inside the overshoot-region).
More realistic spot distributions require the application of a dynamo
model, which determines the spatial and temporal distribution of
magnetic flux at the bottom of the convection zone.
Furthermore, calculated surface distributions for one system cannot
easily be used for other systems with different stellar or orbital
parameters since changes of these quantities alter the tube evolution
and patterns of emergence; individual studies have to be carried out
for each system.

\begin{acknowledgements}
Volkmar Holzwarth thanks Prof.~S.~K.~Solanki and
Prof.~F.~Moreno-Insertis for valuable discussions and the
Max-Planck-Institut f\"ur Aeronomie in Katlenburg-Lindau and the
Kiepenheuer-Institut f\"ur Sonnenphysik in Freiburg/Brsg.~for financial
support during the accomplishment of this work.
\end{acknowledgements}

\end{document}